\documentclass[twocolumn,showpacs,amsmath,amssymb]{revtex4}

\input epsf

\def\Eq#1{Eq.\ (\ref{#1})}
\def\centerbox#1#2{\centerline{\epsfxsize=#1\textwidth \epsfbox{#2}}}

\newcommand{\be}{\begin{eqnarray}}
\newcommand{\ee}{\end{eqnarray}}
\def\st{\begin{equation}}
\def\stp{\end{equation}}
\def\bg{\begin{eqnarray}}
\def\nd{\end{eqnarray}}
\def\commut#1#2{\Big[#1\,,\, #2\Big]}
\newcommand{\non}{\nonumber\\}
 \def\p{{\bf p}}
 \def\k{{\bf k}}

 \def\ansatz{{\it Ansatz}}

\def\alphas{\alpha_{\rm s}}

\def\nf{N_{\rm f}}
\def\mth{m_{\rm th}}
\def\half{\frac{1}{2}}

\usepackage{graphicx}

\begin{document}
\title{Next-to-Leading Order Shear Viscosity in $\lambda \phi^4$ Theory}

\author{Guy D.\ Moore}

\affiliation{Physics Department, McGill University, 3600
rue University, Montr\'eal, QC H3A 2T8, Canada}%

\begin{abstract}
We show that the shear viscosity of $\lambda \phi^4$ theory is sensitive
at next-to-leading order to soft physics, which gives rise to
subleading corrections suppressed by only a half power of the coupling,
$\eta = \frac{(3033.54 + 1548.3 \sqrt{(N+2)\lambda/72}) N T^3}
{\frac{N+2}{3}\lambda^2}$.
The series appears to converge about as well (or badly) as the series
for the pressure.
\end{abstract}


\maketitle

\section{Introduction}

For some time there has been a growing interest in the theoretical
prediction of long-time dynamics and equilibration in hot quantum field
theories.  One driving motivation has been understanding the early
universe, particularly the possibility of electroweak baryogenesis (for
an older but still quite pertinent review see
\cite{RubakovShaposhnikov}).  In this case one is interested in dynamics
of the electroweak sector, SU(2)+Higgs with a coupling $\alpha_{\rm w}
\sim 1/30$.  Another motivation is attempting to describe
high energy heavy ion collisions, which are
being experimentally probed at RHIC and will soon be probed at even
higher energies at the LHC.  Since one of the original motivations for
this program was to explore QCD in a regime where it is thermal and
(relatively) weakly coupled, it makes sense to see how far one can go in
its description using weak-coupling tools, even though the gauge
coupling is not expected to be {\em very} small, $\alphas \sim 1/3$ at
best.

Some of the most interesting and difficult problems involve
understanding long-time dynamics and the approach to equilibrium.
Transport coefficients are a theoretically well-defined and interesting
subset of these problems.  Recently there has been great interest in the
shear viscosity, because RHIC results on elliptic and radial flow and
spectra seem to indicate that the viscosity is surprisingly small
\cite{RHIC_xpt,perfect_hydro,imperfect_hydro}.  On the theoretical side,
great strides have been made in the theoretical evaluation of transport
coefficients.  In a pioneering paper, Jeon showed how to calculate the
shear viscosity of $\lambda \phi^4$ theory at leading order in the
coupling \cite{Jeon}.  He showed that the evaluation reduces to a
problem in kinetic theory.  Since then, kinetic theory treatments of QED
and QCD transport coefficients (electrical conductivity, conserved
number diffusion, and shear and bulk viscosity) have appeared at
leading-log \cite{AMY1} and full leading order \cite{AMY6,Dogan}.  The
kinetic treatment has also been justified diagrammatically at
leading-log \cite{Aarts,someone} and leading order \cite{JSGagnon}.

However, except for two unrealistic simplifying limits
(large $\nf$ gauge theory \cite{largeNF} and large $N$ O($N$) scalar
field theory \cite{AartsResco}), there are no results beyond leading
order in weak coupling for any long-time dynamical quantities in any
interacting relativistic 3+1 dimensional gauge theory%
\footnote{There are beyond-leading order {\em strong coupling} results
    for some highly supersymmetric gauge theories, by means of the AdS-CFT
    correspondence \cite{Buchel}.}.
We feel this is a big hole in the literature, because without any
knowledge of subleading corrections, it is hard to know how quickly
perturbation theory converges.  On the other hand, a subleading result
gives at least a hint of how fast the series converges; when the
subleading correction becomes as large as the leading order result,
perturbation theory has almost surely failed.  With this in mind, this
brief report will compute the shear viscosity of the
simplest toy theory, $N$ component $\lambda \phi^4$ theory at general
$N$, at next-to-leading order.

The reason this is possible, and a main reason it is interesting, is
that the subleading correction emerges from soft physics.  In this
theory, the plasma of scalars induces an effective mass for all scalar
excitations, $\mth^2 = \frac{N+2}{72} \lambda T^2$.  But soft scalars
play an important role as the ``targets'' in scattering processes.  The
scalars which are significantly influenced by this effective mass
represent the target in $O(\mth/T)$ of scattering events. Therefore the
scattering
rate, and with it all long-timescale dynamical quantities, receive
$O(\sqrt{\lambda})$ corrections.  Actually, both this thermal mass and
the mass dependence of the shear viscosity were mentioned by Jeon
\cite{Jeon}, so his 1995 paper technically represented an NLO
calculation, not just the leading-order one he claimed.  But this was
not mentioned (or recognized) in that paper, so we feel that a clear
discussion is warranted.

\section{Leading-order Review}

\begin{figure}[t]
\centerbox{0.35}{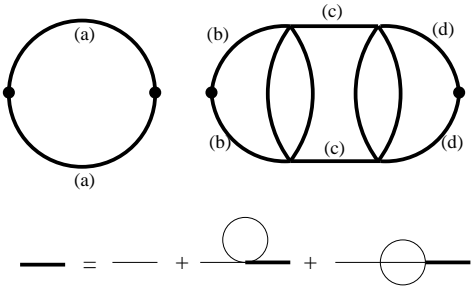}
\caption{\label{fig1} Two leading order diagrams contributing to shear
  viscosity.  Blobs are stress-tensor insertions and heavy lines are
  self-energy resummed as shown.}
\end{figure}

First we review Jeon's leading order calculation, emphasizing
where higher order corrections can arise.  The shear viscosity is
defined as a correction to the (local) equilibrium form for the stress
tensor when a fluid undergoes nonuniform flow.  For a flow velocity
$v_i$ satisfying $\partial_i v_i=0$, the stress tensor in the local rest
frame is
\st
T_{ij} = P \delta_{ij} - \eta \left[ \partial_i v_j + \partial_j v_i
  - \frac{2\delta_{ij}}{3} \partial_k v_k \right] + O(\partial^2)\,,
\stp
with $\eta$ the viscosity.  A fluctuation-dissipation relation (Kubo
formula) relates this coefficient to a certain stress-stress correlation
function \cite{Kubo}:
\st
\eta = \frac{1}{20}\lim_{\k \rightarrow 0} \lim_{\omega \rightarrow 0}
 \int d^4 x e^{i(k_i x_i - \omega x_0)} \; \frac{1}{\omega}
 \left\langle \commut{\pi_{ij}}{\pi_{ij}} \right\rangle \, ,
\stp
where $\pi_{ij} \equiv T_{ij} - \frac{\delta_{ij}}{3} T_{kk}$.
This relation is the starting point for a perturbative analysis.  One
wants the correlation function of two stress operators, at zero spatial
momentum and in the limit of vanishing frequency.  Figure \ref{fig1}
shows the obvious leading-order diagram, together with another diagram
which is also leading-order for a subtle reason.  The heavy lines
represent self-energy resummed propagators.  The propagators with the
same label carry opposite momenta for all values of all loop momentum
integration variables; therefore when one goes on-shell, the other goes
on-shell simultaneously, leading to a pinch singularity.  If not for the
width $\Gamma$, the result would diverge; instead it is enhanced by a
factor of $1 / \Gamma \sim 1/\lambda^2$.  Therefore we must sum all
diagrams with $n$ such pinches and $2n-1$ loops.  More loops without
extra pinches are generally suppressed by a power of $\lambda$ per
loop.

The set of diagrams which we must resum turns out to be all diagrams
structurally like the second diagram of Figure \ref{fig1}, and
up to subleading in $\lambda$ corrections, their resummation
gives precisely the same description of $\eta$ which we would get from
kinetic theory.  The kinetic description takes the plasma to be made up
of weakly coupled quasiparticles which undergo occasional scatterings
described by a Boltzmann equation.  The relevance of quasiparticles
arises because of the importance of nearly on-shell propagators in
evaluating the diagrams; the scattering processes arise from evaluating
the self-energy corrections to these propagators, which give rise to
$\Gamma$.  According to kinetic theory, the stress tensor is determined
by the distribution function for particles $f(x,\p)$ through
\st
T_{ij}(x) = \int \frac{d^3 p}{(2\pi)^3} \frac{p_i p_j}{p^0}
f(x,\p) \,.
\stp
This distribution function evolves according to the Boltzmann equation,
\st
\label{Boltzmann1}
\frac{\partial f(x,\p)}{\partial t} + \frac{p_i}{p^0} 
  \frac{\partial f(x,\p)} {\partial x_i} = -{\cal C}[f] \, ,
\stp
\vspace{-0.2in}
\bg
\label{Boltzmann2}
{\cal C}[f] & = & \half \int \frac{d^3 k d^3 p' d^3 k'}
  {(2\pi)^9 2p^0 2k^0 2p'{}^0 2k'{}^0}
  |{\cal M}_{pk\rightarrow p'k'}|^2 
\non  && \hspace{0.3in} \times
(2\pi)^4 \delta^4(p{+}k{-}p'{-}k') 
\non  && \hspace{0.3in} \times
  \Big( f(\p) f(\k) [1{+}f(\p')][1{+}f(\k')] 
 \non && \hspace{0.5in}
  -  f(\p') f(\k') [1{+}f(\p)][1{+}f(\k)] \Big) \, .
\nd
In writing Bose stimulation functions we have assumed all bosonic
species; we have also neglected higher order scattering processes, as
justified by our expansion in $\lambda$.

Nonuniform flow velocity means that the equilibrium distribution
function would be
$$
f_0(x,\p) = \Big( \exp[ (p^0 - v_i p_i)/(T \sqrt{1-v^2}) ] -1 \Big)^{-1}
$$
and the spatial derivative in \Eq{Boltzmann1} acts on this to give
$(p_i p_j \partial_i v_j) f_0(1{+}f_0)/(p^0 T)$.  This term drives the
system from equilibrium.  The collision term vanishes if $f=f_0$ but is
linear in departures $\delta f = f-f_0$; such a departure will arise and
grow until the collision term and the gradient term cancel (remember
that we want the vanishing frequency or long-time behavior).  The
departure from equilibrium will be of the same structural form as the
spatial derivative term;
\st
f(x,\p) = f_0 + (\partial_i v_j)\frac{p_i p_j}{p^0 T^3} f_0(1{+}f_0)
\chi(|p|)\, ,
\stp
with $\chi(|p|)$ a function of the magnitude of $p$ only, determined by
solving Eqs.\ (\ref{Boltzmann1},\ref{Boltzmann2}).  This is most
conveniently done by defining an inner product $\langle f|g \rangle
\equiv \int \frac{d^3 p}{(2\pi T)^3} f(p) g(p)$ and a functional $Q[\chi]$
\cite{AMY1},
\st
\label{Q}
Q[\chi] = T \langle \chi | {\textstyle \frac{|\p|^2}{p^0}} \rangle
- \frac{1}{2} \langle \chi | {\cal C}_{\rm lin} | \chi \rangle \, ,
\stp
where ${\cal C}_{\rm lin}$ is a linearization of the collision operator
given below.  The $\chi$ which extremizes $Q$ solves the Boltzmann
equation, and the extremal value of $Q$ is $15 \eta/2$.

\section{Subleading corrections}

The advantage of the diagrams is that they help us see where the
perturbative argument that further corrections are down by $O(\lambda)$
can break down.  Specifically, for generic momenta an extra loop is
$O(\lambda)$.  The self-energy resummation is an $O(\lambda)$ correction
unless the propagator momentum is soft or close to the light-cone.  Near
the light-cone, the width correction is essential, but it is already
incorporated in the leading-order calculation and it naively receives
only $O(\lambda)$ corrections from higher order diagrams.  The one-loop
self-energy leads to a mass correction mentioned in the introduction,
$
m^2_{\rm th} = \frac{N+2}{72} \lambda T^2
$
for $N$ component scalar theory with O($N$) symmetry and interaction
$\frac{\lambda}{24} (\phi_n \phi_n)^2$.  This mass correction shifts the
propagating pole by $O(\lambda)$ for generic momentum $p \sim T$.
However, for $p \sim \mth \sim T \sqrt{N\lambda}$, the shift to the
propagating pole is an $O(1)$ correction.  Therefore, in any diagram
where one of the lines is soft, $p \sim \mth$, there will be
$O(1)$ interaction corrections not taken into account in the
leading-order analysis.

How important are soft momenta?  It turns out that $\chi(p)$ vanishes as
$\chi(p) \propto p$ at small momenta, meaning that particles at these
momenta are very near equilibrium and do not contribute significantly to
$T_{ij} - P \delta_{ij}$, despite the $f_0(1{+}f_0)$ enhancement
factor.  Therefore the first term in \Eq{Q} is only sensitive to
dispersion corrections at $O(\lambda)$.
However, such particles {\em are} of some importance as
scatterers.  To see this, first let us estimate what fraction of all
particles are so soft.  The total particle number density is
$n_{\rm tot} = N \int (d^3 p/(2\pi)^3) f_0 \sim N T^3$.  The soft
contribution is 
$n_{\rm soft} \sim N \int (d^3 p/(2\pi)^3) f_0 \Theta(\mth{-}p)
\sim N \mth^2 T \sim (N\lambda) n_{\rm tot}$.  (Phase space gives
three factors of $\mth$ but the distribution function is $f_0 \sim
T/\mth$ in the soft region.)  Soft particles are therefore an
$O(N\lambda)$ fraction of all particles.  In addition, scattering
cross-sections generically scale as $\sigma \sim 1/s$ (the Mandelstam
variable), and when a $k=O(T)$ particle scatters from a 
$p = O(\mth)$ particle, $s \simeq -2p_\mu k^\mu \sim \mth T$ is
smaller than $s \sim T^2$ valid for generic scattering.  Therefore the
cross-section is enhanced by a $T/\mth$ factor, and the relative
rate of scatterings from a soft target is actually $O(\mth/T)
= O(\sqrt{N\lambda})$.

So soft scatterings are an $O(\sqrt{N\lambda})$ fraction of all
scatterings.  Do they matter to equilibration?  The answer is yes; while
the incoming particle is soft, most of the phase space for the
scattering shares the final state energies roughly equally.  This means
that the hard particle retains its direction of flight but substantially
changes its energy, and since the departure from equilibrium has a
strong energy dependence, this is relevant.  To calculate it, we need to
perform the scattering calculation more carefully, incorporating the
soft phase-space dispersion relation.

The collision term we need for the functional $Q[\chi]$ is
\bg
\label{chiCchi}
\langle \chi | {\cal C}_{\rm lin} | \chi \rangle & \!\!\!\equiv \!\!\! &
\int \frac{d^3 p \, d^3 k\, d^3 p'\, d^3 k'}
                {(2\pi)^{12} \: 2p^0 \,2k^0 \,2p'{}^0 \,2k'{}^0 \; T^5}
\\ && \hspace{0.2in} \times
(2\pi)^4 \delta^4(p{+}k{-}p'{-}k')  
\non && \hspace{0.2in} \times
\frac{1}{8} |{\cal M}|^2
f_0(p) f_0(k) [1{+}f_0(p')][1{+}f_0(k')]
\non && \hspace{0.2in} \times
\left( \chi_{ij}(p){+}\chi_{ij}(k){-}\chi_{ij}(p')-\chi_{ij}(k')
\right)^2 \, , \non
\chi_{ij}(p) & \!\!\!\equiv\!\!\! & 
\frac{p_i p_j - \delta_{ij} p^2} {p^0} \; \chi(p) \,.
\nd
The phase space is most conveniently dealt with in terms of the total
energy $\omega = p^0{+}k^0$ and momentum $q = |\p{+}\k|$, two particle
momenta, and an azimuthal angle.  For massless dispersion relations,
this ``s-channel'' parametrization reduces to \cite{largeNF}
\bg
&&\int \frac{d^3 p \, d^3 k \, d^3 p' \, d^3 k'}
                {(2\pi)^{12} \: 2p^0 \,2k^0 \,2p'{}^0 \,2k'{}^0}
(2\pi)^4\delta^4(p{+}k{-}p'{-}k')
\non &=&
\frac{1}{2^8 \pi^5} \int_0^\infty\!\!\!\! d\omega \int_0^\omega\!\!\!\! dq
\int_{\frac{\omega-q}{2}}^{\frac{\omega+q}{2}} \!\!\!\!dp
\int_{\frac{\omega-q}{2}}^{\frac{\omega+q}{2}} \!\!\!dp'
\int_0^{2\pi} \!\frac{d\phi}{2\pi} \,.
\nd
The relations between these variables and the relative angles of the
particles are given in \cite{largeNF}, and can be used to evaluate
\Eq{chiCchi}.

The particle energies are $p$, $p'$, $\omega{-}p$, and $\omega{-}p'$.
The $\omega \sim m$ region is $m^4$ suppressed, as all four variables
must be $\sim m$.  [There is a $(T/m)^4$ enhancement from the population
  functions, but a $(m/T)^4$ suppression from the small value of
  $\chi^2_{ij}$, so this region really is $O(\lambda^2)$.]
A region with one soft particle is $m^2$ suppressed;
for instance, for soft $p$, both $p$ and $\omega{-}q$ must be $O(m)$.  
However, one of the statistical functions in \Eq{chiCchi} is $O(T/m)$ in
this case, and there is no suppression from the values of $\chi_{ij}$.
Therefore the contribution is $O(m/T)$, as stated above.  Note however
that two soft particles (say, $p,p' \sim m$) is $O(m^3)$ suppressed,
compensated by two powers of population functions but further suppressed
by cancellations in the $\chi$ functions; so this region is $O(\lambda)$
at most.

For the massive case it is more convenient to work in terms of particle
energies rather than momenta.  As just discussed, we need only consider
the case where one particle is soft.  In the small $p^0$ region, and
taking $\omega \gg \mth$, the $p^0,q$ phase space is modified by
\st
\int^\omega \!\!\!\!dq \int_{\frac{\omega-q}{2}} \!\!\!\!dp
= \int_0 \!dp \int_{\omega-2p}^\omega \!\!\!\!\! dq 
\;
\Rightarrow \;
\int_m \!\!\!dp^0 \int_{\omega-p^0-p}^{\omega-p^0+p} \!\!dq \,,
\stp
with $p\equiv \sqrt{(p^0)^2-\mth^2}$.
If we consider the difference between the collision integral with and
without massive dispersion, we want the difference between these
integrations.  In this difference region, we can take $p^0$ small,
allowing the substitution $f_0(p^0) = \frac{T}{p^0}$ and
$f(k)=f(\omega)$.  Furthermore, we may neglect $\chi_{ij}(p)$ and set
$\chi_{ij}(k) = (k_i k_j/\omega) \chi(\omega)$.  In this region, $k$ and
the final state momenta $p'$, $k'$ are collinear, so
\bg
&&( \chi_{ij}(p){+}\chi_{ij}(k){-}\chi_{ij}(p'){-}\chi_{ij}(k') )^2
\non & \simeq &
( \omega \chi(\omega) - p' \chi(p') - (\omega{-}p') \chi(\omega{-}p') )^2
\,.
\nd
The $q,p^0$ and $\omega,p'$ integrals then factorize, such that
\bg
&& \langle \chi | {\cal C_{\rm lin}}_{,m=0} | \chi\rangle
- \langle \chi | {\cal C_{\rm lin}}_{,m} | \chi\rangle
\non
&= &
\left[ \int_0 \!\!dp^0 \int_{\omega{-}2p^0}^{\omega} \!\!\!\!dq
\;\frac{T}{p^0} 
-\int_m \!\! dp^0 \int_{\omega-p^0-p}^{\omega-p^0+p} \!\!\!\!dq 
\;\frac{T}{p^0} \right]
\non
& & \times \frac{|{\cal M}|^2}{8} \frac{4}{2^8 \pi^5 T^5}
\int_0^\infty \!\!\!\!d\omega \int_0^\omega \!\!\!\!dp' f_0(\omega)
e^\omega f_0(p') f_0(\omega{-}p')
\non && \hspace{0.3in} \times
( \omega \chi(\omega) - p' \chi(p') - (\omega{-}p') \chi(\omega{-}p')
)^2 \, .
\label{deltaC}
\nd
Here the factor of 4 counts the fact that there are 4 such ``corners''
of phase space where a particle becomes soft, and we have only
considered the contribution of one of them.  Note that the $p'$ integral
is insensitive to the edges $p'\sim 0$ and $p'\sim \omega$ of its range,
so there is no problem of overcounting.  The bracketed integral
gives $\pi \mth T$.


We want to perturb in the small mass.  Therefore it is fair to write the
collision operator as
\st
{\cal C}_{\rm lin} = {\cal C}_{{\rm lin},m=0} + \delta {\cal C}
\stp
with $\delta {\cal C}$ equal to minus the expression in \Eq{deltaC}.
The correction to $Q$ in \Eq{Q} is therefore
$-\half \langle \chi | \delta {\cal C} | \chi \rangle$.  The change in
the extremal value at linear order in $\delta {\cal C}$ is just found by
evaluating this using the unperturbed extremal value of $\chi$.

What remains is to solve the variational problem with the unperturbed
collision operator and to substitute the result for $\chi$ into
\Eq{deltaC}.  We handle the variational problem using the
multi-parameter \ansatz\ method developed in
\cite{AMY1}.  Since we are tuning the function $\chi$ using the $m=0$
collision operator, the evaluation of $\langle \chi|\delta {\cal C}
| \chi \rangle$ only improves linearly with the quality of the
variational \ansatz.  Therefore we must use a large number of the
variational functions presented there; going from 1 to 2 basis functions
changes our result by $4.6\%$, going from 2 to 4 functions changes it
$2.3\%$, going from 4 to 6 functions changes it $0.13\%$, and going from
6 to 8 changes it $0.01\%$.  Our result in the one-component theory,
using an 8 basis function expansion, is
\st
\eta = \frac{3033.54 + 1548.3 \; \mth/T}{\lambda^2} T^3
\stp
where both integration and basis size errors should be smaller than the
last digit shown.  The viscosity increases because the thermal mass
reduces the number of soft targets, and viscosity is inversely related
to the scattering rate.
Jeon found a leading result 5\% smaller and a mass
correction 15\% larger, presumably because of lower quality numerics.
The result can be extended to the O($N$) symmetric theory by scaling the
number of species by $N$ and the scattering rate and thermal mass each
by $(N{+}2)/3$.

\section{Discussion}

We have shown that the first corrections to the leading-order
calculation of the shear viscosity in scalar $\phi^4$ theory arise at
order $O(\sqrt{\lambda})$.
They originate because scattering processes are sensitive to soft
($p \sim T\sqrt{\lambda}$) particles at that order, and soft particles'
behavior is modified by plasma effects at the $O(1)$ level.  All of
these remarks should apply to gauge theories including QCD.  However,
the details of the calculation in those cases will be more complex,
since the dispersion relations are modified in a more complex manner in
gauge theories, and in nonabelian gauge theories the mutual interactions
between soft particles are also strongly modified
\cite{BraatenPisarski}.

One of the motivations for this work was to analyze the convergence of
the perturbative treatment.  What we have shown is that the first
subleading correction is a small correction provided 
$\mth \ll 2 T$.  This is satisfied if $\lambda \ll \frac{288}{N{+}2}$.
The thermal pressure converges in a comparable sized range
\cite{Taylor}.  So does vacuum perturbation theory; the ratio of the
$\beta$ function to the coupling
$\frac{\beta_\lambda}{\lambda}=\frac{N{+}8}{96\pi^2}\lambda$
is small for $\lambda \ll \frac{96\pi^2}{N{+}8}$.
Therefore the
subleading corrections only become large when the theory is approaching
its Landau pole.
Since the origin of the leading-order
correction is well understood, we might hope that the explicit inclusion
of the thermal mass in the external state dispersion, and a resummation
of the thermal mass via a gap equation, might improve the convergence of
the series, as has been argued for the calculation of the pressure
\cite{Petreczky}.  Aarts and Martinez-Resco have done this in the large
$N$ theory, where it is well justified at leading order in a $1/N$
expansion \cite{AartsResco}.

\centerline{\bf Acknowledgements}

This work was supported in part by 
the Natural Sciences and Engineering Research Council of Canada,
and by le Fonds Nature et Technologies du Qu\'ebec.


\begin{thebibliography}{99}

\bibitem{RubakovShaposhnikov}
V.~A.~Rubakov and M.~E.~Shaposhnikov,
  Usp.\ Fiz.\ Nauk {\bf 166}, 493 (1996)
  [Phys.\ Usp.\  {\bf 39}, 461 (1996)]
  [hep-ph/9603208].

\bibitem{RHIC_xpt} See for instance,
K.~Adcox {\it et al.}  [PHENIX Collaboration],
  Nucl.\ Phys.\  A {\bf 757}, 184 (2005)
  [nucl-ex/0410003];
J.~Adams {\it et al.}  [STAR Collaboration],
  Nucl.\ Phys.\  A {\bf 757}, 102 (2005)
  [nucl-ex/0501009].

\bibitem{perfect_hydro} See for instance,
D.~Teaney, J.~Lauret and E.~V.~Shuryak,
  nucl-th/0110037;
P.~Huovinen, P.~F.~Kolb, U.~W.~Heinz, P.~V.~Ruuskanen and S.~A.~Voloshin,
  Phys.\ Lett.\  B {\bf 503}, 58 (2001)
  [hep-ph/0101136].

\bibitem{imperfect_hydro}
P.~Romatschke and U.~Romatschke,
  arXiv:0706.1522 [nucl-th].

\bibitem{Jeon}
S.~Jeon,
  Phys.\ Rev.\  D {\bf 52}, 3591 (1995)
  [hep-ph/9409250].


\bibitem{AMY1}
P.~Arnold, G.~D.~Moore and L.~G.~Yaffe,
  JHEP {\bf 0011}, 001 (2000)
  [hep-ph/0010177].

\bibitem{AMY6}
P.~Arnold, G.~D.~Moore and L.~G.~Yaffe,
  JHEP {\bf 0305}, 051 (2003)
  [hep-ph/0302165].

\bibitem{Dogan}
P.~Arnold, C.~Dogan and G.~D.~Moore,
  Phys.\ Rev.\  D {\bf 74}, 085021 (2006)
  [hep-ph/0608012].


\bibitem{Aarts}
G.~Aarts and J.~M.~Martinez Resco,
  JHEP {\bf 0211}, 022 (2002)
  [hep-ph/0209048];
  Phys.\ Rev.\  D {\bf 68}, 085009 (2003)
  [hep-ph/0303216].

\bibitem{someone}
M.~A.~Valle Basagoiti,
  Phys.\ Rev.\  D {\bf 66}, 045005 (2002)
  [hep-ph/0204334].

\bibitem{JSGagnon}
J.~S.~Gagnon and S.~Jeon,
  Phys.\ Rev.\  D {\bf 75}, 025014 (2007)
  [hep-ph/0610235].

\bibitem{largeNF}
G.~D.~Moore,
  JHEP {\bf 0105}, 039 (2001)
  [hep-ph/0104121].

\bibitem{AartsResco}
G.~Aarts and J.~M.~Martinez Resco,
  JHEP {\bf 0402}, 061 (2004)
  [hep-ph/0402192].

\bibitem{Buchel}
A.~Buchel, J.~T.~Liu and A.~O.~Starinets,
Nucl.\ Phys.\ B {\bf 707}, 56 (2005).

\bibitem{Kubo}
R.~Kubo,
  J.\ Phys.\ Soc.\ Jap.\  {\bf 12}, 570 (1957).


\bibitem{BraatenPisarski}
E.~Braaten and R.~D.~Pisarski,
  Nucl.\ Phys.\  B {\bf 337}, 569 (1990).

\bibitem{Taylor}
J.~Frenkel, A.~V.~Saa and J.~C.~Taylor,
  Phys.\ Rev.\  D {\bf 46}, 3670 (1992).

\bibitem{Petreczky}
F.~Karsch, A.~Patkos and P.~Petreczky,
  hep-ph/9708244.

\end{thebibliography}
\end{document}